\documentclass[5p,number]{elsarticle}

\usepackage{lineno}
\usepackage{amssymb}
\usepackage{gensymb}
\usepackage[colorlinks=true,linkcolor=blue,urlcolor=blue,citecolor=blue]{hyperref}
\usepackage{amsmath}

\journal{Journal of \LaTeX\ Templates}

\graphicspath{{figures/}}

\begin{document}

\begin{frontmatter}

\title{Off-line Commissioning of the St. Benedict Radiofrequency Quadrupole Cooler-Buncher}

\author[ND,Argonne]{D.P.~Burdette}
\author[ND]{R.~Zite}
\author[ND]{M.~Brodeur\corref{cor1}}\ead{mbrodeur@nd.edu}
\author[ND,Argonne,Manitoba]{A.A.~Valverde}
\author[ND]{O.~Bruce}
\author[ND]{R.~Bualuan}
\author[ND]{A.~Cannon}
\author[Argonne]{J.A.~Clark}
\author[ND]{C.~Davis \corref{cor2}}
\author[ND]{T.~Florenzo}
\author[LLNL]{A.T.~Gallant}
\author[ND]{J.~Harkin}
\author[ND]{A.M.~Houff}
\author[ND,Jiaotong]{J.~Li}
\author[ND]{B.~Liu}
\author[ND]{J.~Long}
\author[ND]{P.D.~O'Malley}
\author[ND]{W.S.~Porter}
\author[ND]{C.~Quick}
\author[FRIB]{R.~Ringle}
\author[ND]{F.~Rivero}
\author[Argonne]{G.~Savard}
\author[ND]{A.~Yeck}

\address[ND]{Department of Physics and Astronomy, University of Notre Dame, Notre Dame, IN, USA}
\address[Argonne]{Physics Divison, Argonne National Laboratory, Argonne, IL, USA}
\address[LLNL]{Nuclear and Chemical Sciences Division, Lawrence Livermore National Laboratory, Livermore, California, USA}
\address[Jiaotong]{School of Physics, Xi'an Jiaotong University, Xi'an, Shaanxi, China}
\address[FRIB]{Facility of Rare Isotope Beams, 
Michigan State University, East Lansing, MI, USA}
\address[Manitoba]{Department of Physics and Astronomy, University of Manitoba, Winnipeg, Manitoba, Canada}

\cortext[cor2]{Current affiliation: Department of Physics and Astronomy, University of Pennsylvania, Philadelphia, PA, USA}
\cortext[cor1]{Corresponding author}

\begin{abstract}
The St.~Benedict ion trapping system, which aims to measure the $\beta-\nu$ angular correlation parameter in superallowed-mixed mirror transitions, is under construction at the University of Notre Dame. These measurements will provide much-needed data to improve the accuracy of the $V_{ud}$ element of the CKM matrix. One of the major components of this system is the radio frequency quadrupole cooler-buncher, which is necessary to create low-emittance ion bunches for injection into the measurement Paul trap. The off-line commissioning of the cooler-buncher, using a potassium ion source, determined that the device could produce cooled ion bunches characterized by a 50-ns full-width-half-maximum time width at its exit. The commissioning results also determined the trapping efficiency to be 93(1)$\%$ and the trapping half-life to be 20.0(5) s.  
\end{abstract}


\end{frontmatter}

%\linenumbers

\section{Introduction}

Precision experimental studies of nuclear $\beta$ decay using ion traps rely on specialty devices to prepare the ideal beam properties for the desired measurement. This is the case for the \textbf{S}uperallowed \textbf{T}ransition \textbf{Be}ta \textbf{Ne}utrino \textbf{D}ecay \textbf{I}on \textbf{C}oincidence \textbf{T}rap (St.~Benedict) \cite{Brodeur2016-StBenedict, OMalley2020-StBenedict,Porter2023-StBenedict} which is currently under construction at the University of Notre Dame's Nuclear Science Laboratory (NSL). St.~Benedict is part of a larger experimental program at the NSL that focuses on superallowed $T=1/2$ nuclear $\beta$ decay between mirror nuclei, expanding upon previously performed precision half-life measurements \cite{17F,25Al,11C,15O,29P,13N} of such decays. These transitions are particularly interesting because they can be used to extract the $V_{ud}$ element of the CKM matrix \cite{Towner2010-Vud}. However, this extraction requires knowledge of the Fermi-to-Gamow-Teller mixing ratio $\rho$ \cite{Naviliat2009-Mirror}. As such, St.~Benedict will be used to measure the $\beta-\nu$ angular correlation parameter, $a_{\beta \nu}$, from which $\rho$ can be derived. 

However, several critical beam preparation steps are required before a measurement using the radioactive ions produced by the NSL \textit{TwinSol} facility \cite{Becchetti2003-TwinSol} can be made in the Paul trap \cite{Brodeur2023-StBenedict}. First, the fast \textit{TwinSol} beam needs to be stopped, transported, and extracted from a large volume gas catcher \cite{Savard-2003-GasCell}. Then it needs to be guided, using a Radio Frequency (RF) carpet \cite{Davis2022-Static} and a Radio Frequency Quadrupole (RFQ) ion guide, through multiple differentially pumped pressure regions \cite{Davis2022-Flow} before injection into the RFQ cooler-buncher \cite{Valverde2019-RFQCB}, which will form ion bunches that are then transported to the measurement Paul trap \cite{Porter2025-PaulTrap}.

The St.~Benedict RFQ cooler-buncher plays the crucial role of converting the continuous beam from the gas catcher and extraction system to a low-emittance bunched beam for efficient injection into the measurement Paul trap. This particular cooler-buncher use the mechanical design of the cooler-buncher that provides bunches to the Facility for Rare Isotope Beam's Electron Beam Ion Trap (EBIT) \cite{Lapierre-2018-NSCLEBIT}. That design has also been employed for cooler-bunchers at the $N=126$ Factory \cite{Valverde2020} and CARIBU's \cite{caribu} low-energy stopped beam experimental area at Argonne National Laboratory.

In this paper\footnote{This publication is partly based on a PhD thesis \cite{Burdette-thesis} and includes additional measurements.}, we present the various components of the St. Benedict cooler-buncher as well as the results of its off-line commissioning using a potassium ion source. This commissioning focused primarily on studying the impact of operating pressure, holding time, and various electrostatic and dynamic potentials on the transport efficiency and the full width half maximum (FWHM) of the temporal structure of the ion bunches exiting the device. We aim to maximize efficiency and minimize the FWHM.        

\section{The St. Benedict cooler-buncher}

\begin{figure}
	\centering
	\includegraphics[width=\linewidth]{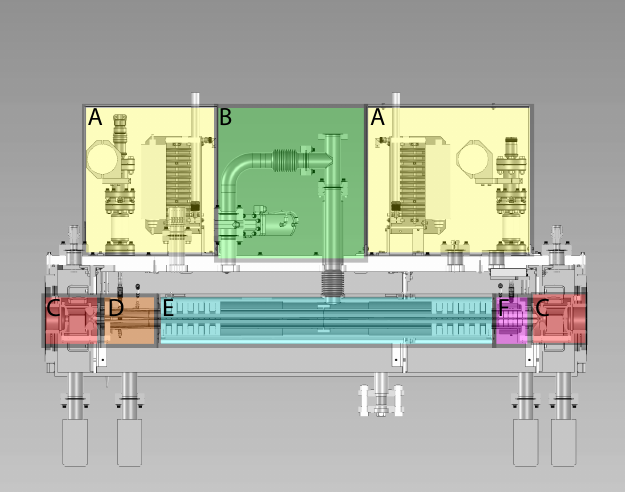}
	\caption{CAD drawing depicting the cross section of the St.~Benedict cooler-buncher. Sections A, in yellow, show the RF circuitry mounted to the lid of the cooler-buncher chamber. Separate circuits for the cooling and bunching sections allow for the application of different RF amplitudes and frequencies to each section. Section B, in green, shows the tower feeding gas into the RFQ cooling region with a bypass to the outer region. Section C, in red, shows the injection/ejection optics. Section D, in orange, shows the flared RFQ rods. Section E, in blue, shows the cooling region. The 11 light-colored tooth pattern rectangles of both ends of the section E represent the PEEK baffles holding the helium and creating the pressure differential. Section F, in magenta, shows the bunching region.}
	\label{fig::CADcoolerBuncher_highlight}
\end{figure}

The RFQ Cooler-Buncher employed at St. Benedict, shown in Fig. \ref{fig::CADcoolerBuncher_highlight}, follows the same working principles as any typical gas-filled RFQ \cite{werth2009charged}. It will take the $\sim$200 eV energy beam from the differentially pumped gas catcher extraction system \cite{Davis2022-Flow} and cool it by collisions with helium atoms at a pressure in the $\sim$10$^{-2}$ Torr range. These thermalized ions will then be dragged through the gas volume using an electric field produced by a decreasing potential applied on a segmented quadrupole structure down to the bottom of a potential well where they will accumulate. The transported ions are confined radially by a harmonic potential created by the application of a time-varying quadrupolar potential at RF on four electrodes encircling the beam. After some time, the accumulation of ions in the cooler-buncher is interrupted and the captured ions are released by rapidly changing the potential on electrodes in proximity to the potential well. 

The St.~Benedict cooler-buncher includes some of the advanced features implemented in the BECOLA cooler-buncher \cite{BARQUEST-2018} to improve the acceptance and efficiency of the RFQ while being able to handle a higher beam current. These features, together with the overall characteristics of the St.~Benedict cooler-buncher will be described in the following sections. 

\subsection{RFQ injection and ejection optics}

\begin{figure}
	\centering
	\includegraphics[width=\linewidth]{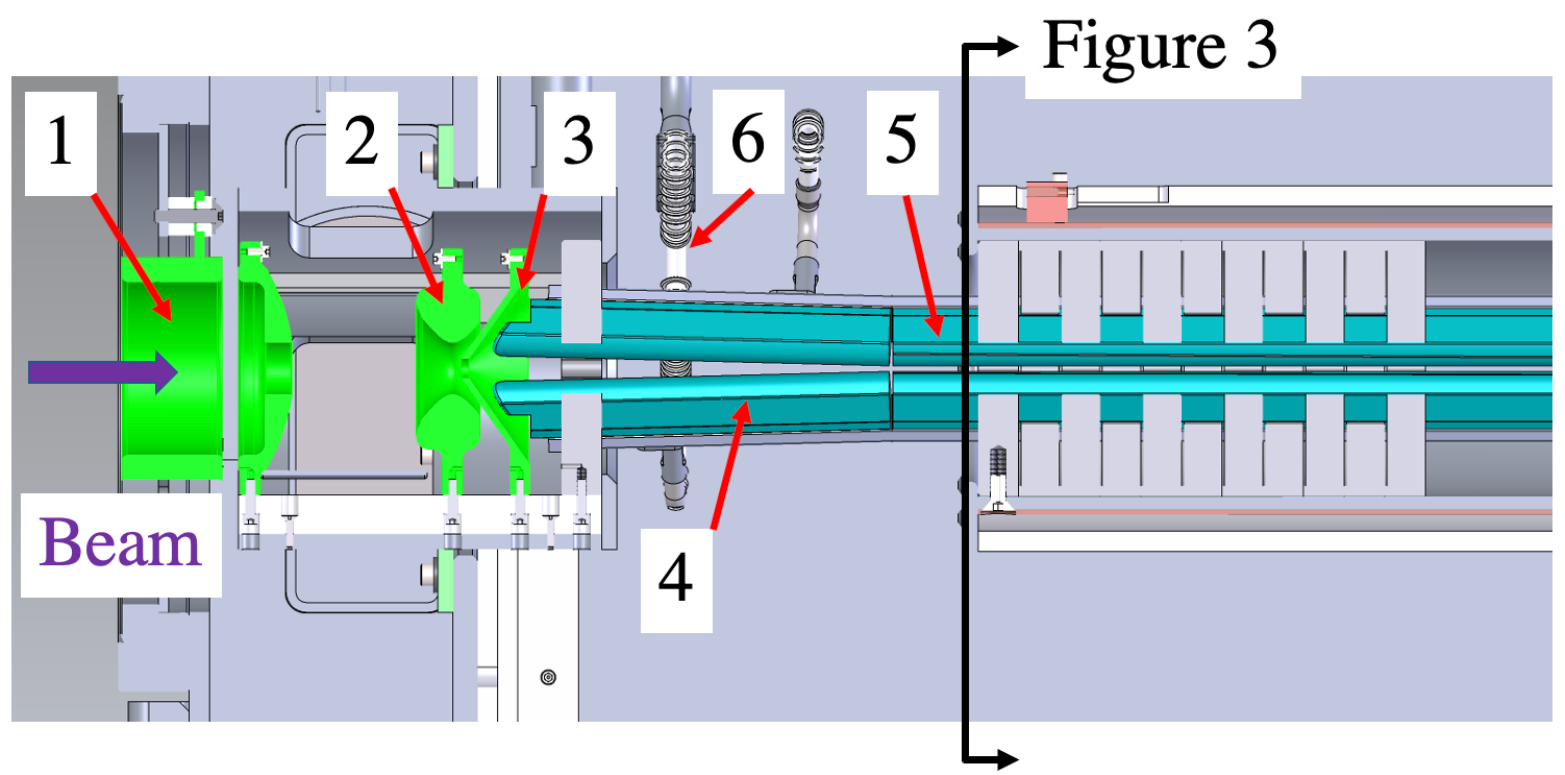}
	\caption{A zoomed-in view of the upstream side of the RFQ cooler-buncher. The three injection electrodes, namely the 1) lens, 2) hyperbola, and 3) cone, are highlighted here in green. The 4) flared RFQ segment A and the 5) RFQ segment B and C are in cyan. The 6) Stethoscopic RF connection is also marked. The location of the cross-section cut of Fig.~\ref{fig::crosscutinjection} is indicated here as well.}
\label{fig::injectionOptics}
\end{figure}

The upstream side of the RFQ (labeled C on the left side of Fig.~\ref{fig::CADcoolerBuncher_highlight}) is designed to maximize the acceptance of the ion beam. Incoming ions first encounter 3 injection electrodes, shown in Fig.~\ref{fig::injectionOptics} and are called, in order, the injection lens, hyperbola, and cone electrodes. At the appropriate electric potential, these electrodes provide the necessary focusing that maximizes acceptance into the RFQ. Similarly, the ejection side of the RFQ (labeled C on the right side of Fig.~\ref{fig::CADcoolerBuncher_highlight}) also contains, in order, an ejection cone, hyperbola, and lens electrode aimed at efficiently focusing the bunches exiting the cooler-buncher.  

To further aid in the acceptance of ions on the upstream side of the RFQ, the first set of RFQ rods (labeled D in Fig.~\ref{fig::CADcoolerBuncher_highlight}) are flared away from the beam axis, at an optimal $2^\circ$ opening angle \cite{BARQUEST-2018} to further direct off-axis ions into the RFQ (see Fig.~\ref{fig::injectionOptics}). The electrodes then taper back towards the axis so that the downstream side matches the 3.5-mm inscribed radius of all the other RFQ electrodes.

\subsection{Cooling section}\label{sec::Coolingsection}

\begin{figure}
	\centering
	\includegraphics[width=\linewidth]{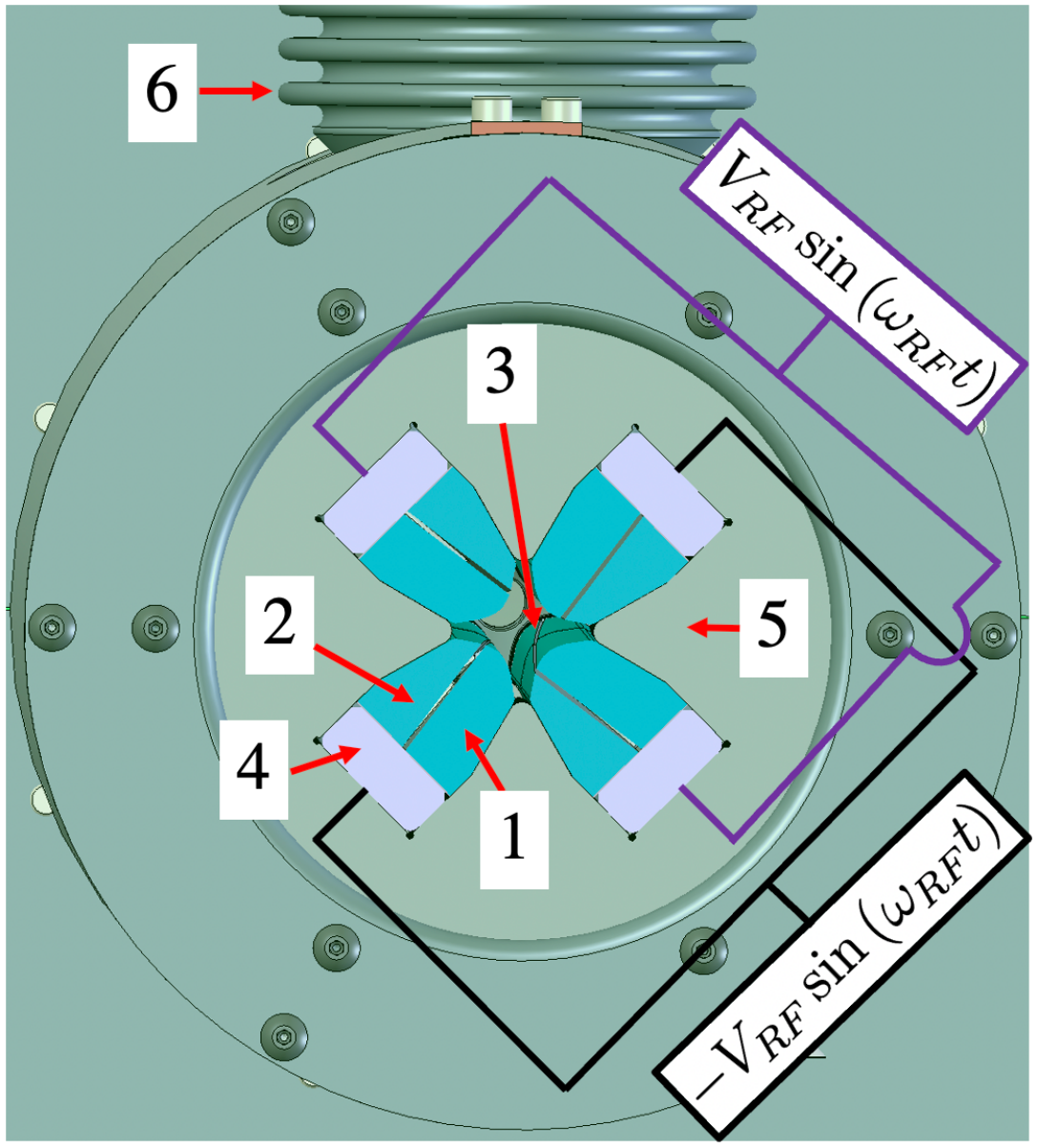}
	\caption{Cross-cut of Fig.~\ref{fig::injectionOptics} with a slight tilt showing electrodes 1) B and 2) C, as well as the 3) turning cross-cut between them and the 4) RF backbone. The application of the opposite RF phase is also shown schematically. The 5) PEEK piece serving as differential-pumping barrier enveloped the electrode structure. Helium gas is injected through the 6) bellow above the cooling section.}
\label{fig::crosscutinjection}
\end{figure}

Another feature of this RFQ is its separate pressure regions for the cooling and bunching of the ion beam, labeled E and F, respectively, in Fig.~\ref{fig::CADcoolerBuncher_highlight}. This separation is accomplished by a physical sheath around the cooling region. At both ends of the cooling region, the rods pass through a set of cut-out PEEK baffles used to contain the helium, as shown in Fig.~\ref{fig::crosscutinjection}. The gas is therefore forced to evacuate through on-axis openings on both ends. The outside regions adjacent to this sheath, which also house the flared RFQ electrode on the upstream side (D in Fig.~\ref{fig::CADcoolerBuncher_highlight}) and the bunching electrodes on the downstream side (F in Fig.~\ref{fig::CADcoolerBuncher_highlight}), are pumped by one turbo molecular pump each.  There is another barrier to further separate the injection and ejection electrodes, and these regions (C in Fig.~\ref{fig::CADcoolerBuncher_highlight}) are pumped by a turbo molecular pump each.  The gas is then continuously flowed into the cooling region and the various turbo molecular pumps maintain the lower pressures outside of the sheath. These different pressure regions allow for the cooling of ions as they traverse the cooling region, while minimizing the loss of ions directly upstream and downstream.

\subsection{Electrode geometry}

\begin{figure}
	\centering
	\includegraphics[width=\linewidth]{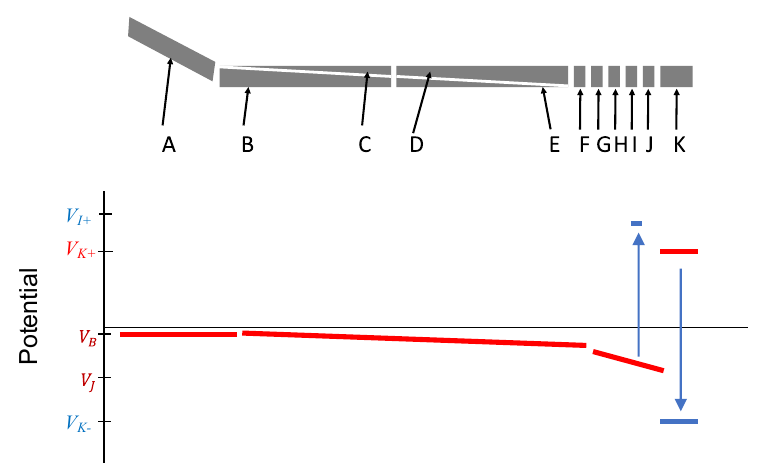}
	\caption{The potential scheme for cooling and accumulating ions in the bunching section, shown in red, and the potential scheme for ejecting an ion bunch from the bunching region, in blue and indicated by the arrows. The two schemes are implemented by changing potentials on electrodes I and K, as labeled above.}
	\label{fig::potentials}
\end{figure}

\begin{figure}
	\centering
	\includegraphics[width=\linewidth]{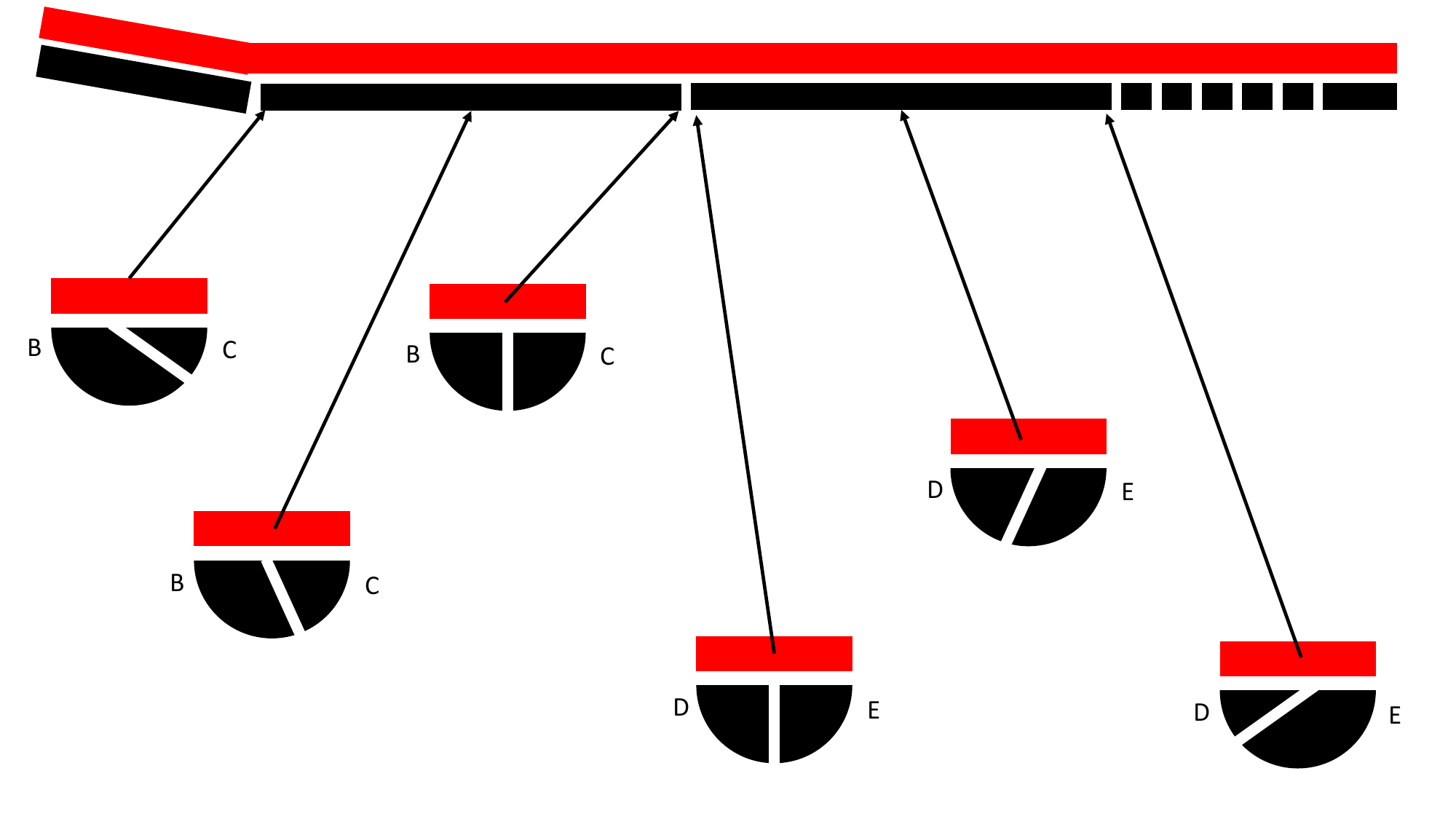}
	\caption[Schematic of cross-cut electrodes in the cooling section of the St. Benedict RFQ cooler-buncher.]{A rough schematic showing cross sectional views of the electrodes of the cooling region depicting the segmentation in six different spots to create a smooth drop in potential along the beam axis.}
	\label{fig::rfq_coolingElectrodes}
\end{figure}

All electrodes of the RFQ, including those of the flared, cooling and bunching region, have a hyperbolic cross section defining the surface facing the center of the RFQ with a flat back on the opposing side that is capacitively coupled to a separate rectangular electrode (referred to as the backbone electrode) directly behind it, as shown in Fig.~\ref{fig::crosscutinjection}. The constant potential (DC) for each hyperbolic electrode is applied directly to the RFQ electrode through holes in the backbone. RF is applied to the so-called backbone (see Fig.~\ref{fig::crosscutinjection}), and is capacitively coupled to the individual electrodes via a thin Kapton sheet. This allows one RF signal to supply identically phased RF to all electrodes of each segment while maintaining the ability to apply independent DC voltages to each segment along the beam axis. For each rod, there are a total of 11 segments along the beam axis which are referenced alphabetically, with the first labeled electrode A and the last electrode K. A depiction of rod segmentation is given in Fig.~\ref{fig::potentials} and \ref{fig::rfq_coolingElectrodes}. There are two types of segmentation for this RFQ, including the cooling section comprised of cross-cut electrodes B through E (see Fig.~\ref{fig::crosscutinjection}) and a bunching section comprising electrodes F through K.

The two types of rod segmentation, comprising the cooling region and the bunching region, offer different advantages for the two different goals they are intended to achieve.  After ions pass through flared electrode A they enter the cooling region of electrodes B, C, D, and E.  A rough schematic showing the cross section of these electrodes at six different points along the beam axis is shown in Fig.~\ref{fig::rfq_coolingElectrodes}. At the front of this region, the potential is dominated by electrode B, since there is a negligible cross-section of electrode C facing the center of the RFQ.  As the ions traverse the length of electrode B, they will see a linear change in potential from the average of the potentials on electrodes B and C because of the diagonal cut on these electrodes. The same behavior continues along the lengths of electrodes D and E. However, the potential at the front of this region is denoted by the average of electrodes D and E, and the end of the region is defined mainly by the potential on electrode E. These two linear regions provide both a smoother drag potential and fewer individual electrical connections than the simple segmentation scheme found in the bunching section. 

However, the structure of the bunching section is more ideal for the accumulation of ions in a certain region. The bunching region is made up of five identical length electrodes (F-J) and one longer electrode (K). To allow for the accumulation and subsequent extraction of ion bunches, high-voltage switches are used for electrodes I and K. To accumulate ions, voltages are applied in such a manner as to create a linear drop in potential from electrodes F to J and a positive voltage on electrode K (see Fig.~\ref{fig::potentials}). This creates an ion cloud between electrodes I and K. Then, to extract ions, the voltage on electrode I is pulsed up, and the voltage on electrode K is pulsed down. The cycle of accumulating and ejecting ions is then repeated (at a typical repetition rate of 20 Hz) so that the continuous DC beam is converted into a bunched beam. 

\subsection{RF circuitry}

\begin{figure}
	\centering
	\includegraphics[width=\linewidth]{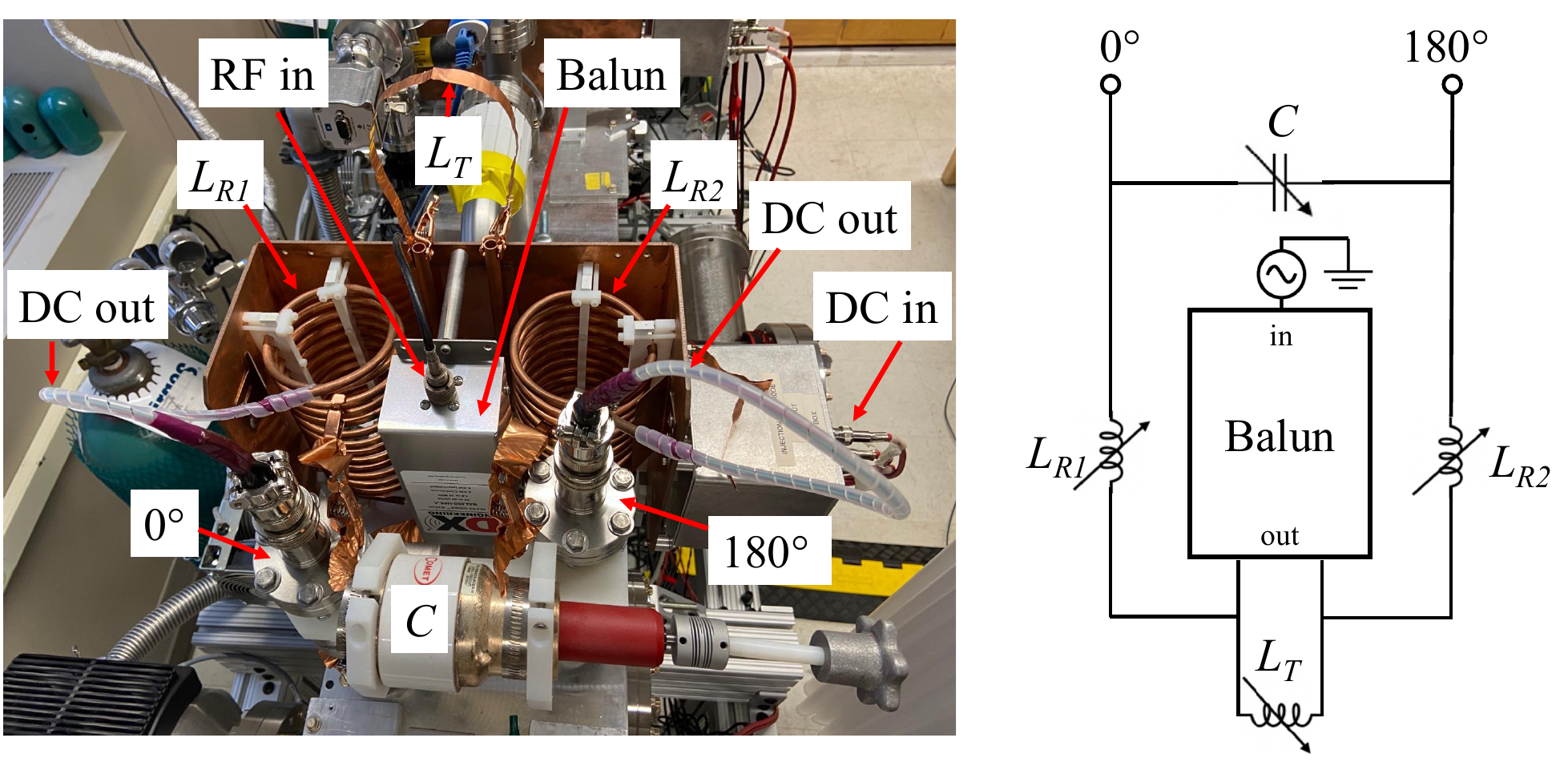}
	\caption{Left: Photo of the injection-side RF circuit and DC connections to the feedthroughs. Right: RF circuit diagram. The main circuit components such as the RF input, balun, large variable coils ($L_{R1}$ and $L_{R2}$), the variable impedance matching inductor ($L_T$), the variable capacitor $C$, and the feedthroughs receiving the 0$\degree$ and 180$\degree$ phases are indicated. The left photo also show the DC input and breakout box as well as the multicore cables outputting from the large variable coils.}
	\label{fig::RFcircuit}
\end{figure}

The RF circuit has been designed to provide a signal of large amplitude, on the order of 1000 Volts peak-to-peak, on the various electrodes and was based on the BECOLA cooler-buncher circuit \cite{BARQUEST-2018}. Fig.~\ref{fig::RFcircuit} shows the RF circuit that first includes a DX Engineering MAXI-Core Balun transformer model BAL050-H05-A to create the two opposing-phase RF signals sent to the electrodes. After the Balun, the two RF phase signals are sent to separate coils made of 0.19" ID copper tubes that serve as the inductive component of the resonant LC circuit. A variable capacitor is placed between the two opposite ends of these coils to allow adjustment of the circuit's resonant frequency. The coils are attached to electrically isolated feedthroughs that transmit the RF signal, under vacuum, through stethoscope-shaped stainless steel tubes that connect the feedthrough to the RF backbones as shown in Fig. \ref{fig::injectionOptics}. The ends of the coils are also connected through a so-called trombone inductor that is adjusted to match the low impedance of the resonant LC circuit to the 50$\Omega$ impedance of the RF generator. Half-inch-wide copper strips and copper alligator clips were used to connect these various components, allowing maximum freedom in circuit tuning, including independently adjusting the inductance of the two coils labeled $L_{R1}$ and $L_{R2}$ in Fig.~\ref{fig::RFcircuit} 
while minimizing resistive losses.

RF is provided to the cooler-buncher electrodes via two different RF circuits for maximum flexibility. The injection RF circuit (indicated by the left A region in Fig.~\ref{fig::CADcoolerBuncher_highlight}) covers electrodes A to C. This includes the first half of the cooling section. The extraction RF circuit (indicated by the right A region in Fig.~\ref{fig::CADcoolerBuncher_highlight}) covers electrodes D to K. This includes the second half of the cooling section and the bunching section. The two circuits are operated at sufficiently different resonant frequencies of 4.98 MHz for the injection, and 3.66 MHz for the extraction, such that they are outside of each other bandwidth to avoid cross-amplification of their signals. These frequencies are in the range of cooler-buncher devices of similar design \cite{Lapierre-2018-NSCLEBIT,BARQUEST-2018}.
The RF signal is provided by two T$\&$C AG1006 300 W RF generators.

\subsection{DC connections}

The DC potential for all electrodes is provided by W-Ie-Ne-R EHS MPOD high voltage modules and fed to the breakout boxes shown in Fig.~\ref{fig::RFcircuit}. A 230 nF polyfilm capacitor connected to ground was added to all DC segments in the breakout box for which the potential is not switched to maintain a stable potential that is unaffected by the switching of the other electrodes. The DC connections exit the breakout box in a multicore unshielded cable that is then fed inside the inductive copper coil and attached to a multipin connector on the opposite end that is connected to the RF feedthrough. Once under vacuum, the DC connections are fed inside the stethoscope-shaped connectors (indicated in Fig. \ref{fig::injectionOptics}) on which the RF is applied. All of these features, in concert with the RF backbone, maximize the capacitive coupling between the RF and DC.

The potentials on electrodes I and K (see Fig.~\ref{fig::potentials}) are rapidly switched to open and close the trap using two switches from Argonne National Laboratory originally manufactured by McGill University that are rated to $\pm250 V$. A 270 nF polyfilm capacitor connected to ground has been added to the high and low inputs of the switches to help the EHS power supplies recover faster to their set voltages after switching. The measured 10$\%$ to 90$\%$ rise and fall time of the switches is around 200 ns. Finally, the only other electrode that experiences a rapidly changing potential is the injection lens electrode, which serves as the beam gate. The potential on this electrode is rapidly switched between ground and 300 V, using a 3 kV Belhke switch, to allow the incoming beam to be blocked. 

\subsection{Timings}

\begin{figure}
	\centering
	\includegraphics[width=\linewidth]{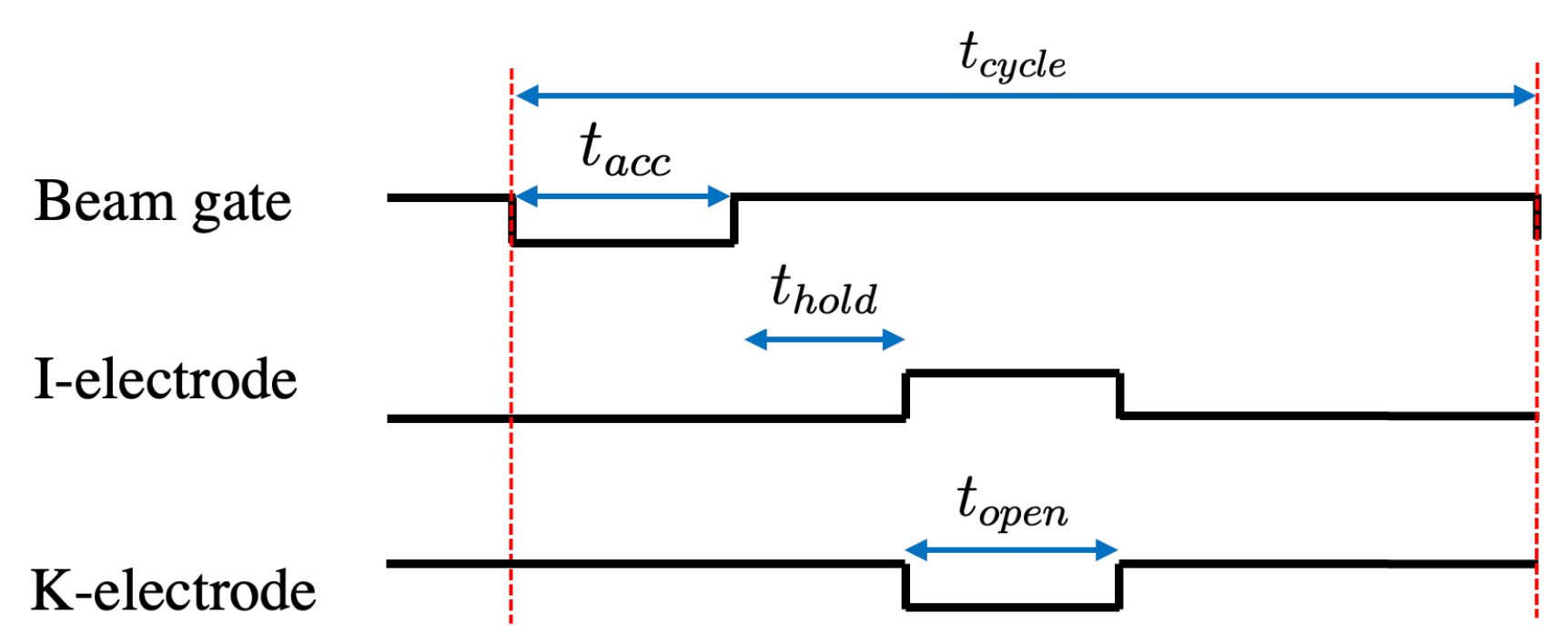}
	\caption{Timing sequence used in bunching mode. Cycle duration $t_{cycle}$ is indicated by the red doted lines. The cycle starts with the opening of the beam gate and accumulation of ions for a duration $t_{acc}$. The ions are then held in the bunching section for a duration $t_{hold}$. Finally they are released by lowering (raising)  the potential on electrode K (I) for a duration $t_{open}$. The duration of the various timings are not up to scale.}
\label{fig::TimeSequence}
\end{figure}

The trigger signals for the various switches were provided by an SRS DG535 digital delay and pulse generator. Fig.~\ref{fig::TimeSequence} shows a typical timing sequence when operating in bunch mode. For the various results presented, a 50 ms cycle time was used. Each cycle started with the reduction of the injection lens potential for a duration, $t_{acc}$, during which the beam accumulates in the RFQ. The injection lens potential is then raised to block the beam and the ions are held in the bunching section for a duration $t_{hold}$. Then the potential on electrode K is rapidly lowered, while the potential is simultaneously raised on electrode I to eject the ion bunch. The trap is kept open for a duration $t_{open}$ before being closed again in preparation for the next cycle. 

For the presented studies, an accumulation time of 10 $\mu s$ was typically used, as it provided a sufficient number of ions while avoiding pileup on the microchannel plate (MCP) detector. The ions were also held in the RFQ for 40 ms (unless otherwise noted) and the trap was open for 200 $\mu s$ to ensure that all ions had exited before it was closed again. Such an extended timing pulse can also help extend the lifetime of the switches.

\subsection{Gas handling and pumping system}

\begin{figure}
	\centering
	\includegraphics[width=\linewidth]{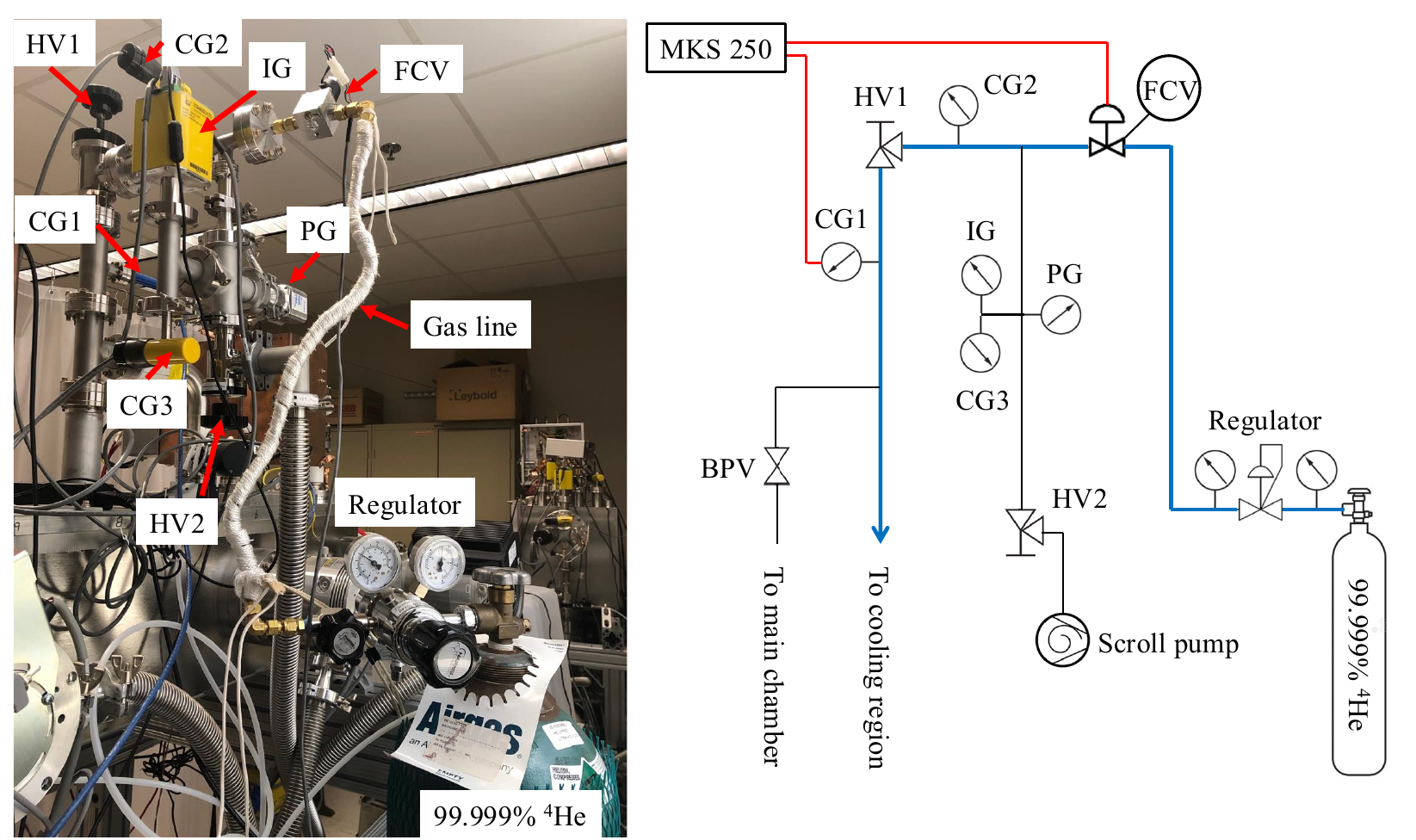}
	\caption{Left: Photo of the gas handling system for the RFQ cooler-buncher. Right: Corresponding schematic diagram. The thick blue line indicates the flow path taken by the gas in normal operation with hand valve HV1 open and HV2 closed. In this operation the flow control valve FCV is controlled by the MKS250 unit that takes the pressure from the convectron gauge CG1 (electrical connections between these components denoted in red). The piezo gauge PG, the convectron gauge CG2 and CG3 and the ion gauge IG are used to display gas line pressure from atmosphere to high vacuum. While the bypass valve BPV to the main chamber and the hand valve HV2 to the scroll pump are indicated on the schematic, they are not visible on the photo.}
\label{fig::gashandling}
\end{figure}

Ultra-high purity (99.999$\%$) helium was used for the measurements presented. As indicated in Fig.~\ref{fig::gashandling} the gas bottle was connected to a copper line that was cleaned according to UHV standards and baked at 120$\degree$C for 12 hours. That line is connected to a flow control valve controlled by an MKS flow control unit that uses a KJ Lesker Pirani gauge located along the gas feed line for reference. The flow control valve is connected to a vacuum enclosure that can be isolated from the RFQ cooler and buncher via a hand valve. That enclosure can also be evacuated using a scroll pump via a line normally isolated by a second hand valve. Finally, that enclosure includes the various pressure gauges indicated in Fig.~\ref{fig::gashandling}.
%an MKS piezo gauge for an absolute pressure measurement when the helium pressure is above 0.1 Torr, a Worker Bee convection gauge and a Hornet ion gauge both from InstruTech for sub-0.1 Torr pressure measurements. The enclosure is connected to the ``gas feed tower'' (labeled B in Fig.~\ref{fig::CADcoolerBuncher_highlight}) which connects directly to the tube containing the cooling region. That tower also includes the KJ Lesker Pirani gauge used in conjunction with the gas flow controller. 

\begin{table}
	\centering
	\caption{Measured pressures using a pirani gauge in each section of the RFQ cooler-buncher with and without flowing helium.}
	\begin{tabular}{ c c c }
		\hline
		\hline
		Section & Without (Torr) & With  (Torr) \\
		\hline 
		Injection optics  & 4.1$\times$10$^{-8}$ & 1.7$\times$10$^{-5}$\\
		Flared RFQ & 9.1$\times$10$^{-8}$ & 9.5$\times$10$^{-5}$ \\
		Bunching & 8.6$\times$10$^{-8}$ & 1.0$\times$10$^{-4}$ \\
		Ejection optics  & 2.0$\times$10$^{-8}$ & 1.2$\times$10$^{-5}$ \\
		Gas feed tower & 6.4$\times$10$^{-6}$ & 6.3$\times$10$^{-2}$ \\
		\hline 
		\hline
	\end{tabular}
	
	\begin{tabular}{l}
	\end{tabular}
	\label{tab::pressures}
\end{table}

The helium is evacuated by four turbo molecular pumps backed by a single Ecodry 65 pump from Leybold. The volumes immediately adjacent to the cooling sections (labeled D and F in Fig.~\ref{fig::CADcoolerBuncher_highlight}) are pumped by 260 L/s TwisTorr turbo molecular pumps from Agilent. The volumes housing the injection and ejection optics (labeled C in Fig.~\ref{fig::CADcoolerBuncher_highlight}) are pumped by 350 L/s turbo molecular pumps from Leybold. Table~\ref{tab::pressures} gives typical pressures, both without and with helium flowing (at a typical rate) in each of the main pumping volume where vacuum gauges are installed. The injection optics chamber pressure, with helium flow, of 1.7$\times$10$^{-5}$ Torr is very close to the anticipated pressure of 1.5$\times$10$^{-5}$ Torr \cite{Davis2022-Flow} in that chamber once the system is on-line and a 75 Torr pressure is used in the gas catcher. As a result, similar levels of 200 eV ion beam losses due to the pressure level can be anticipated once on-line.

\section{Off-line commissioning setup}

\begin{figure}
	\includegraphics[width=\linewidth]{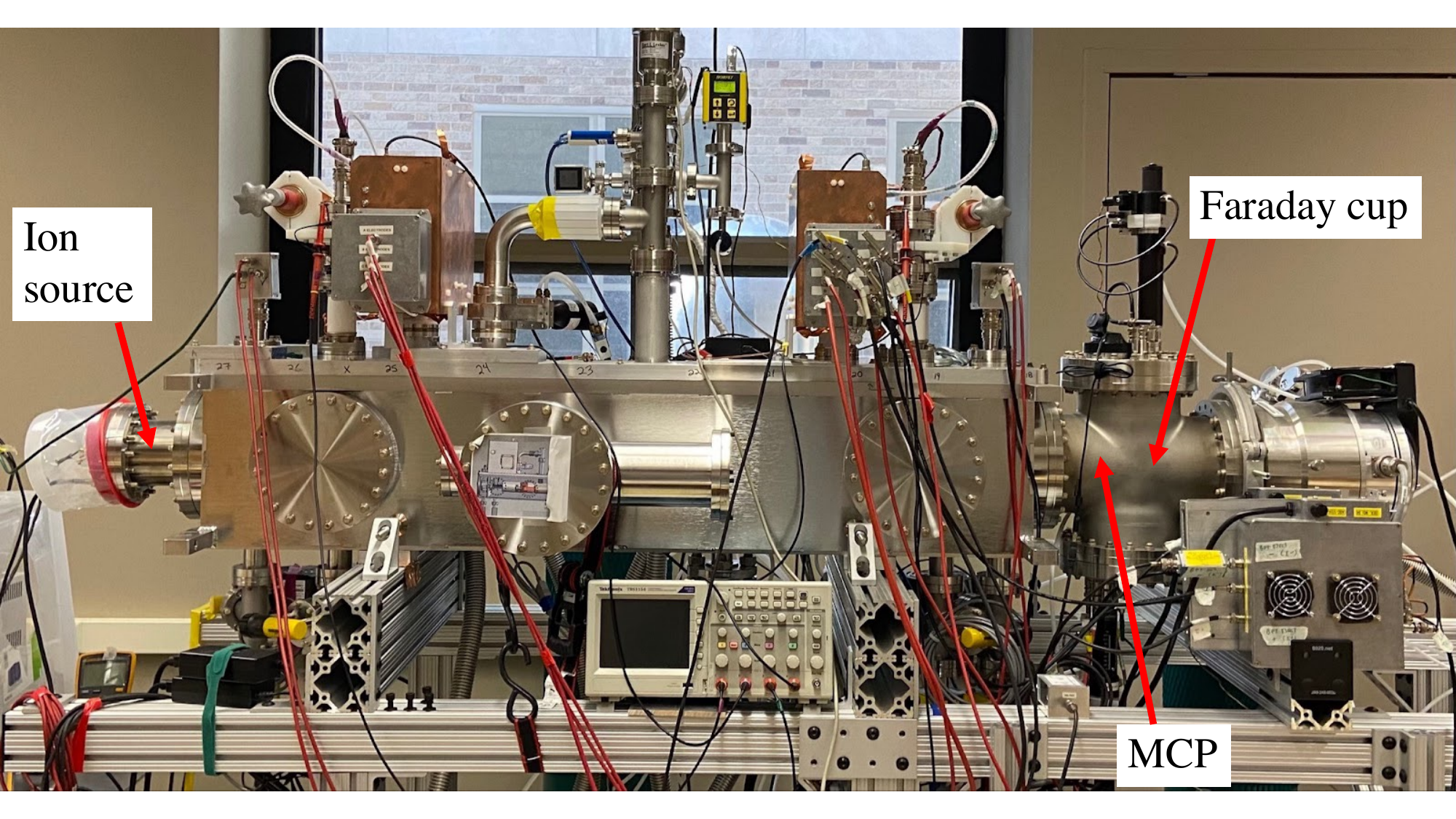}
	\caption{The commissioning setup of the RFQ cooler-buncher. The RFQ cooler-buncher is located inside the large rectangular chamber, as it will be in the St.~Benedict beam line, while a ion source is mounted to the upstream side of the chamber and a 8" conflat cross housing diagnostic detectors is mounted downstream.}
	\label{fig::rfq_setup}
\end{figure} 

To perform the off-line commissioning of the St. Benedict RFQ cooler-buncher, an ion source was installed on the injection side of the device. On the extraction side, we installed a four-way cross with a Faraday cup and MCP as a beam diagnostic. To reduce the helium pressure at the location of the MCP, this cross was pumped with a 550 L/s turbo molecular pump from Agilent backed by an IDP-15 dry scroll vacuum pump also from Agilent. A photo of the test setup is shown in Fig.~\ref{fig::rfq_setup}.

\begin{figure}
	\includegraphics[width=\linewidth]{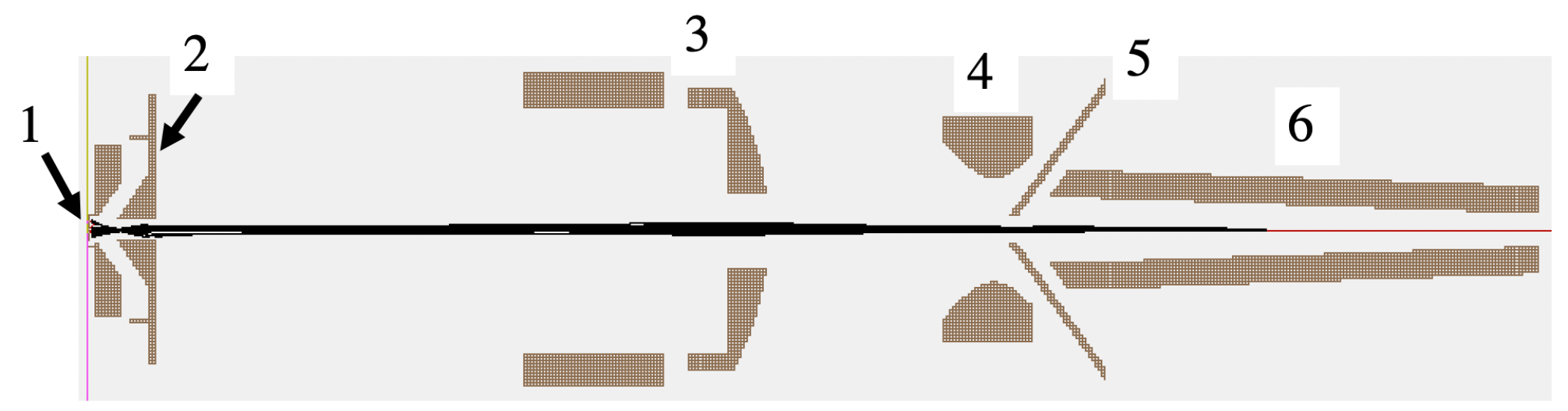}
	\caption{SIMION simulation of the injection of $^{39}$K$^+$ ions produced by the ion source when 200 V and 193.5 V are applied on the ion source and anode respectively. Indicated features include the 1) ion source, 2) anode, 3) injection lens, 4) injection hyperbola, 5) injection cone, and 6) electrode A.}
	\label{fig::sim-injec}
\end{figure} 

For commissioning, we used a potassium thermionic ion source from Heat Wave Labs \cite{heatwave}. For the production of ions, a current of around 1.52 A was typically used. The source, along with its holder, floated to a potential of 200~V. Directly facing the ion source is the anode electrode that sits at a potential of 193.5 V. This ideal extraction potential was found from ion optical simulations (see Fig.~\ref{fig::sim-injec}). The 200 eV beam from the source then passes through the lens electrode maintained at 0 V before being focused using the injection cone and hyperbola electrodes, both at -1400 V, at the entrance of the RFQ cooler-buncher. 
 
The MCP detector used for these measurements was manufactured by Beam Imaging Solutions and includes a phosphor screen used mainly for initial beam tuning. The MCP front plate, back plate, and anode were biased, respectively, at -2000V, 0V, and 200V throughout our measurements for efficient detection while minimizing background counts. 

We determined the bunch profile as a time-of-flight spectrum by feeding the MCP signal into an SRS SR430 multichannel scaler that received the same trigger signal as provided by the SRS DG535 pulser to the switches for the ion bunch extraction. The various settings were systematically varied to optimize the bunch by maximizing the number of counts in the bunch and minimizing the time width of the bunch. Due to the large parameter space to explore, this process was automated by implementing a Labview code which could scan the experimental parameters and record time of flight spectra for a set number of ejections at each setting. After each scan the user evaluates the results and decides on the subsequent parameter(s) to scan. In this way, the parameter space could be explored, one or two parameters at a time, to optimize the parameters and see if they are mutually dependent.  

\section{Commissioning Results}

In this section, an overview of parameter optimization and identified systematic trends is outlined. The beam was first optimized without bunching to maximize transport through the RFQ and get a ``rough'' tune. Then, with the bunched beam, the parameter space for all voltage, timing, pressure, and RF settings was explored to maximize the number of ions extracted while minimizing the time-of-flight spread of the bunches.

\subsection{Injection and ejection optics scan}

\begin{figure}
	\includegraphics[width=\linewidth]{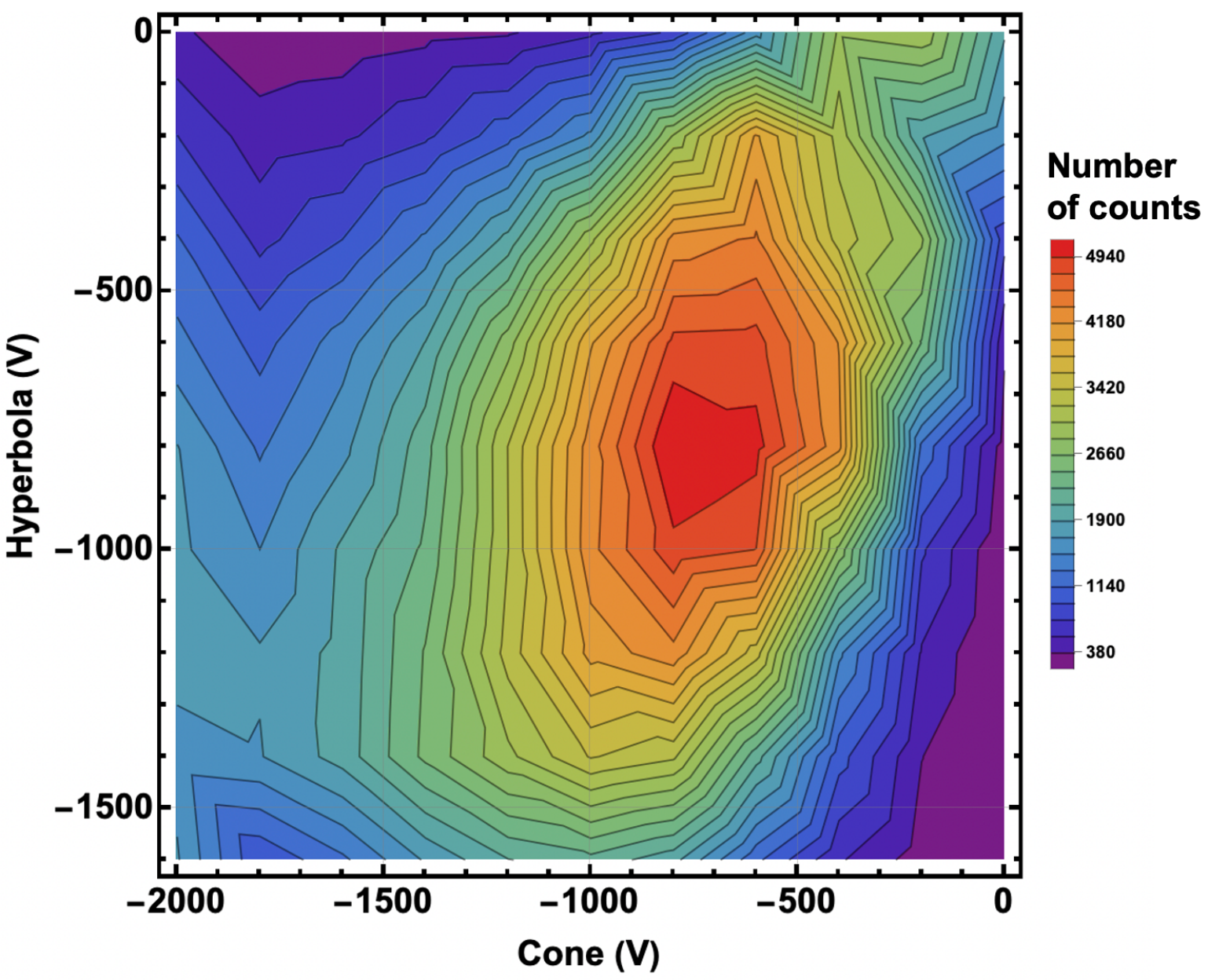}
	\caption{Contour plot of the total number of detected ions exiting the RFQ cooler-buncher for 2000 cycles of a two dimensional scan of the potential applied on the injection hyperbola and cone electrode.}
	\label{fig::voltageScan}
\end{figure}

After tuning the continuous beam, the seven electrodes that comprise the injection and ejection optics were optimized in bunch mode. Potentials were scanned, two at a time, by accumulating thousands of bunches at a repetition rate of 20 Hz for multiple combinations of settings. As an example, Fig.~\ref{fig::voltageScan} shows the total number of recorded ions for a typical two-dimensional scan of the injection hyperbola and cone electrode potential. This figure indicates optimal injection cone and hyperbola potentials of around -750 V and -800 V respectively, resulting in around 5000 ions recorded (for 2000 cycles). This process was repeated for various combinations of all the applied potentials, and repeated multiple times after resetting various voltages to random values to try to get as close as possible to the global maximum of transmission independent of any state the system was in. The same process was repeated for all operating parameters of the cooler-buncher, including all electrode potentials, RF amplitude, pressure, and timing.

\subsection{Operating pressure}

Proper cooling of the ions is critical in the bunching process. To do so, the cooling region needs to be maintained at a sufficiently high pressure to thermalize the ions.

\begin{figure}
	\includegraphics[width=\linewidth]{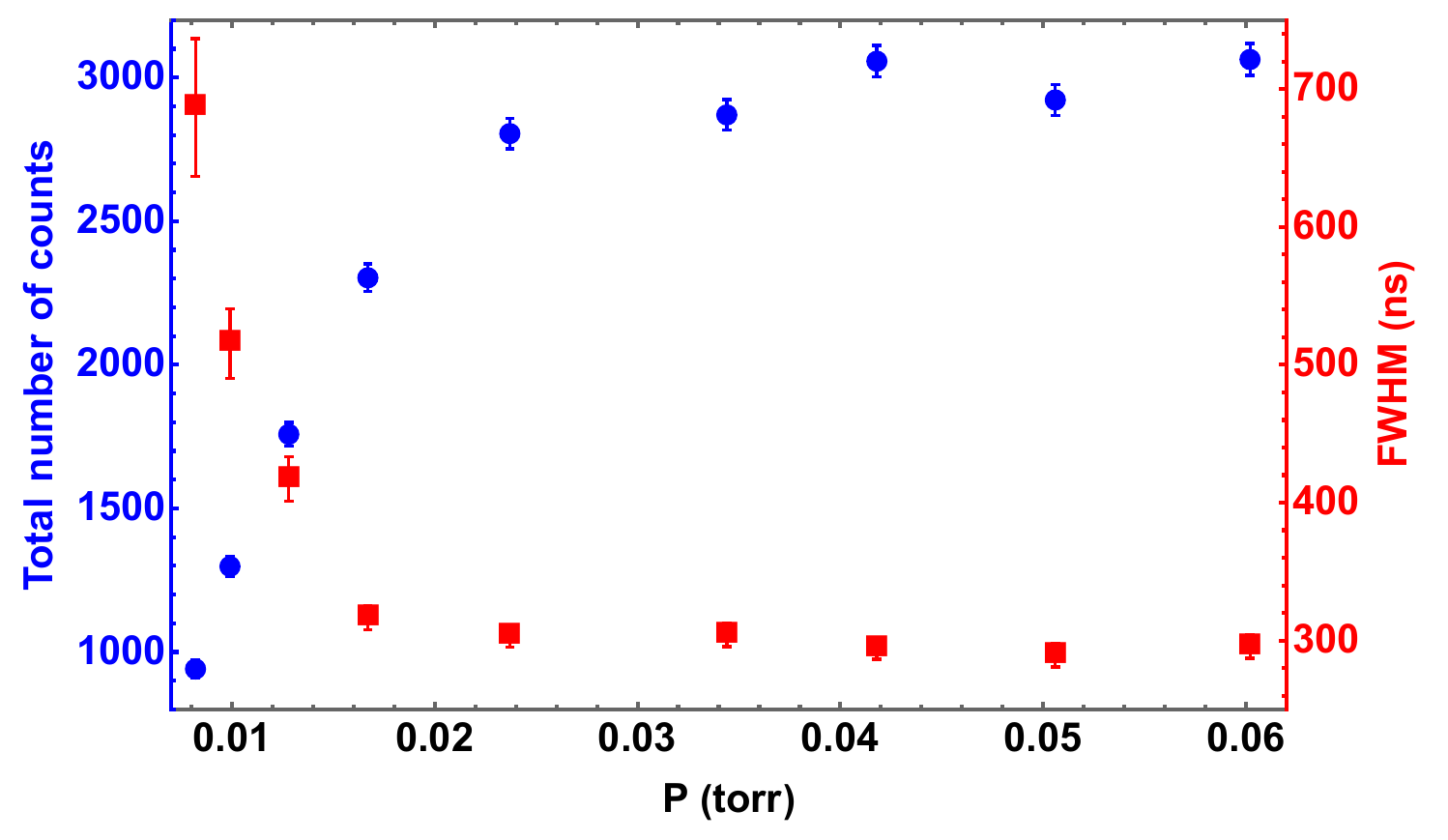}
	\caption{Recorded ion counts (circle) and fitted FWHM (square) of the $^{39}$K$^{+}$ time-of-flight distribution for different pressures.}
	\label{fig::PvsNtotFWHM}
\end{figure} 

To investigate the minimal pressure required to cool the ions, the cooler-buncher was filled with ions for 10 $\mu$s, then the ions were accumulated and held in the bunching region for 40 ms before being released toward an MCP by only lowering the potential on electrode K (electrode I was not simultaneously raised to ensure that no ions remain trapped upstream when the bunch is released). The fitted FWHM of the time-of-flight distribution as well as the total number of ions contained in these bunches as a function of pressure are shown in Fig.~\ref{fig::PvsNtotFWHM}. As can be seen, the FWHM of the bunch rapidly drops before plateauing above 0.02 Torr (as read by a Pirani gauge along the gas line). The number of ions in the bunch displays a similar behavior, indicating that a pressure of at least 0.02 Torr is required to sufficiently cool the ions. 

\subsection{Drag field and holding time}

The effect of the electric field dragging the ions through the cooling section (drag field), on the transport efficiency, for the 50 ms cycle used in this study, has been investigated by sending 10 $\mu$s bursts of ions through the RFQ without capturing them. The drag field is determined by the relative DC electric potentials of the B, C, D, and E electrodes of the RFQ.

\begin{figure}
\centering
	\includegraphics[width=0.95\linewidth]{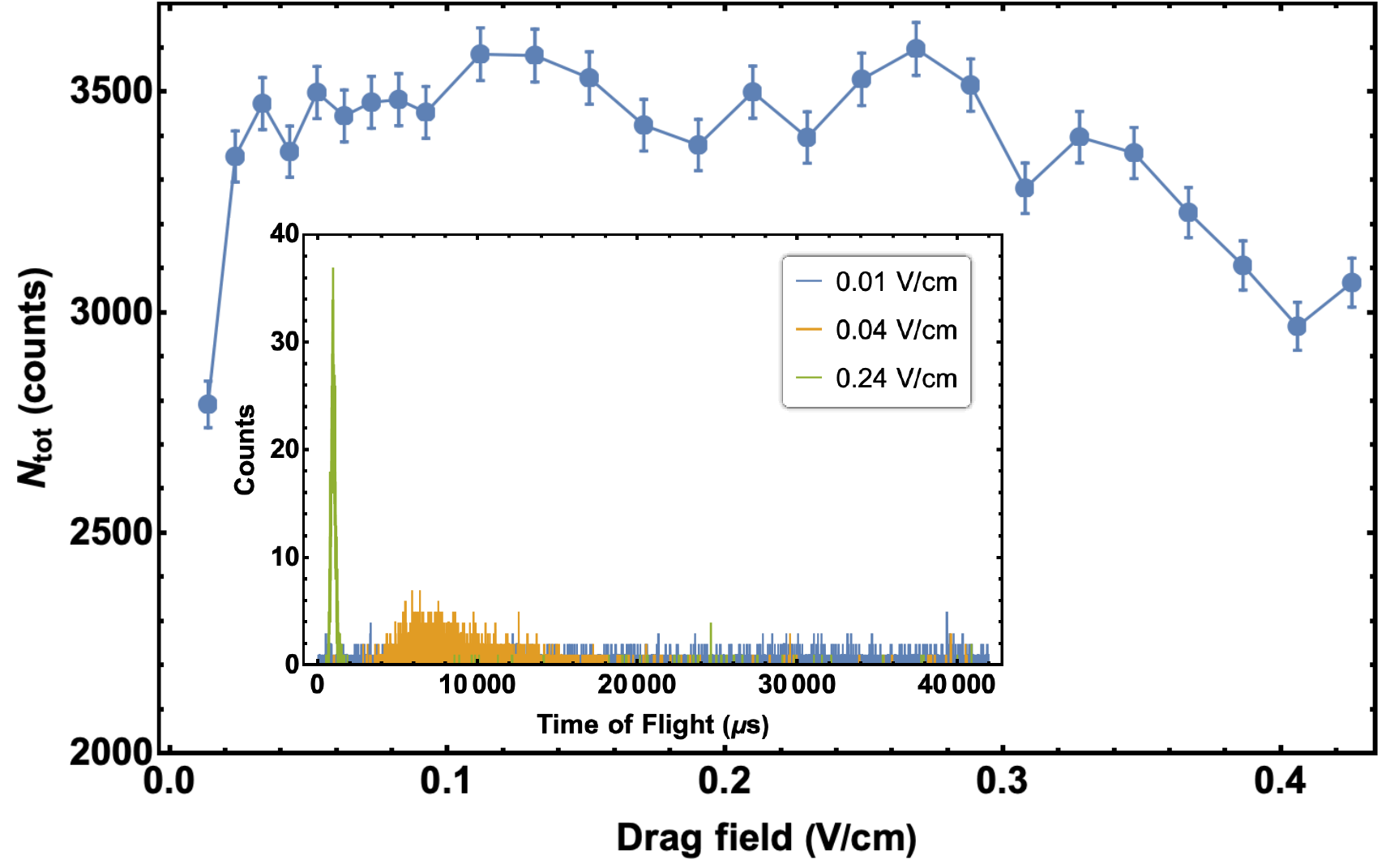}
	\caption{Number of ions $N_{tot}$ recorded within a 42 ms window after the creation of the 10$\mu$s burst for different drag fields. Inset shows the TOF profile of the recorded ions for three different drag fields.}
	\label{fig::EdragNtot}
\end{figure} 

Fig.~\ref{fig::EdragNtot} shows that with increased drag field, the time-of-flight profile of the transported ion burst gets narrower. At a very low drag field the TOF spread of the burst exceeds the 50 ms duration of the cycles used in bunch mode. This means that once in bunching mode, not all ions in a given burst will accumulate in the trapping region before the opening of the trap. However, from 0.03 V/cm and onward the whole TOF fits within the cycle duration, and minimal changes in the number of transported ions are observed for drag fields of up to 0.3 V/cm.

Although most of the cooling occurs in the cooling section, a small amount of it also needs to happen in the bunching section for the ions to accumulate and settle at the bottom of the potential well located at electrode J. Consequently, the ions need to be held in the bunching section a minimum amount of time before being released.

\begin{figure}
\centering
	\includegraphics[width=\linewidth]{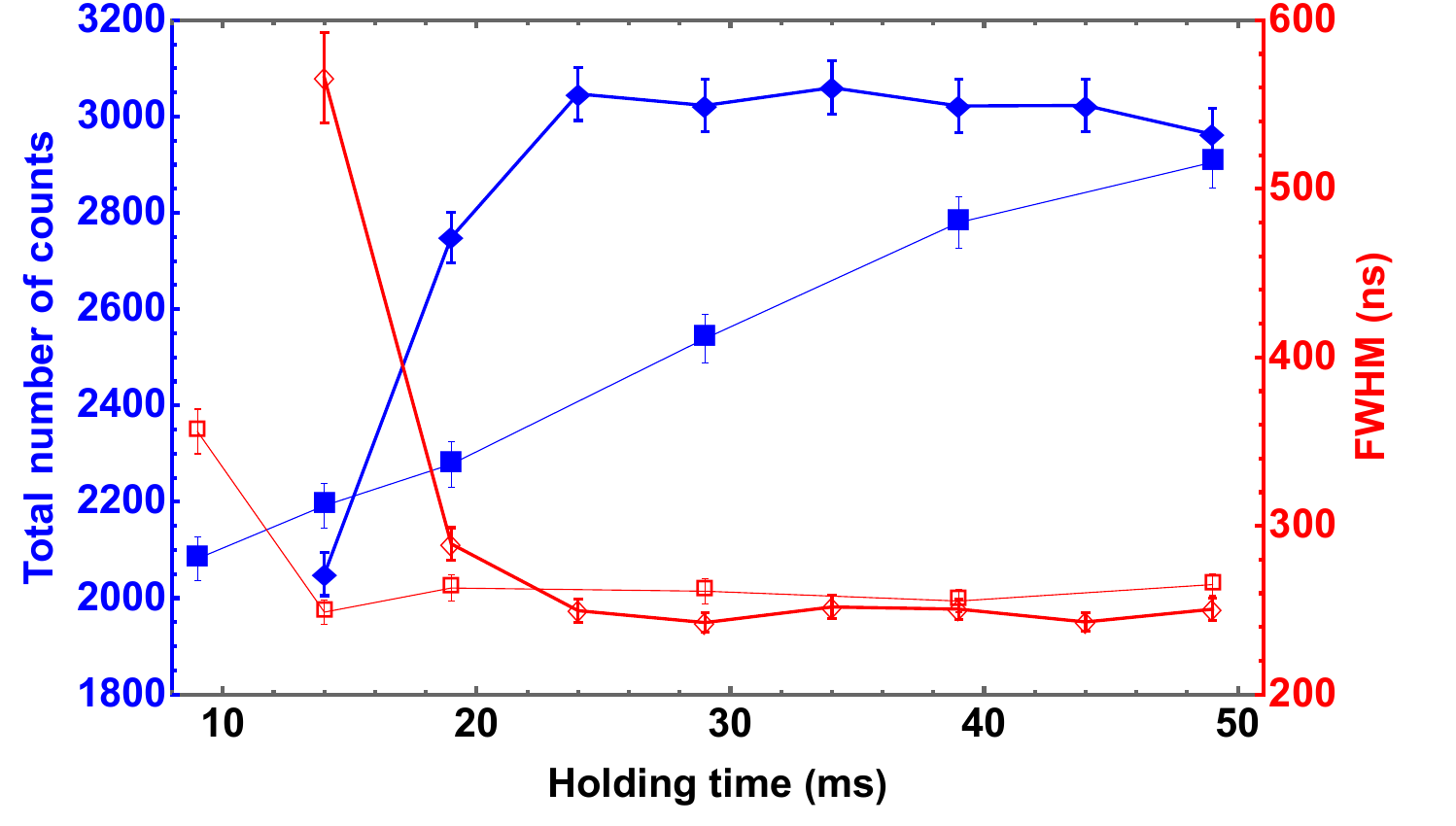}
	\caption{Total number of extracted ions (solid blue) and FWHM (open red) of the $^{39}$K$^{+}$ bunch for different holding times. Results are shown for two different drag fields, indicated by the shape of the data point. Diamond points indicate results from a drag field of 0.04~V/cm and square points for a drag field of 0.24~V/cm.}
\label{fig::tholdvsNtotFWHM-Edrag}
\end{figure} 

The effect of this holding time, $t_{hold}$, has been studied for two different drag fields. To ensure that all ions are released after each opening, we kept the trap open until near the end of the cycle, and only electrode K was lowered to release the ions. If electrode I is simultaneously raised, there is a chance that insufficiently cooled ions remain trapped upstream of electrode I upon extraction.  Fig.~\ref{fig::tholdvsNtotFWHM-Edrag} shows that for drag fields of 0.04 V/cm the number of ions rapidly increases until a holding time of 24 ms after which it plateaus. Similarly, behavior is seen for the decreasing FWHM. This is an indication that the ion pulse has not cooled completely for lower holding times (see the TOF profile in Fig.~\ref{fig::EdragNtot}). A 0.24 V/cm drag field results in more ions being released at small holding time due to the shorter TOF needed for the pulse to traverse the cooling region (see the TOF profile in Fig.~\ref{fig::EdragNtot}). However, the number of ions released increase more gradually with holding time and remain lower than for the 0.04 V/cm drag field. This could be due to an insufficient cooling in the cooling section (see Sec.~\ref{sec::Coolingsection}), requiring extra cooling in the bunching region, which is enabled by a longer holding time.   

\subsection{Potentials in the trapping region}

The bunching section potentials that maximize the number of ions in a bunch were optimized by adjusting the potential slope bridging the region between the end of the cooling section and the trap, as well as the depth of the trap.  

\begin{figure}
\centering
	\includegraphics[width=0.95\linewidth]{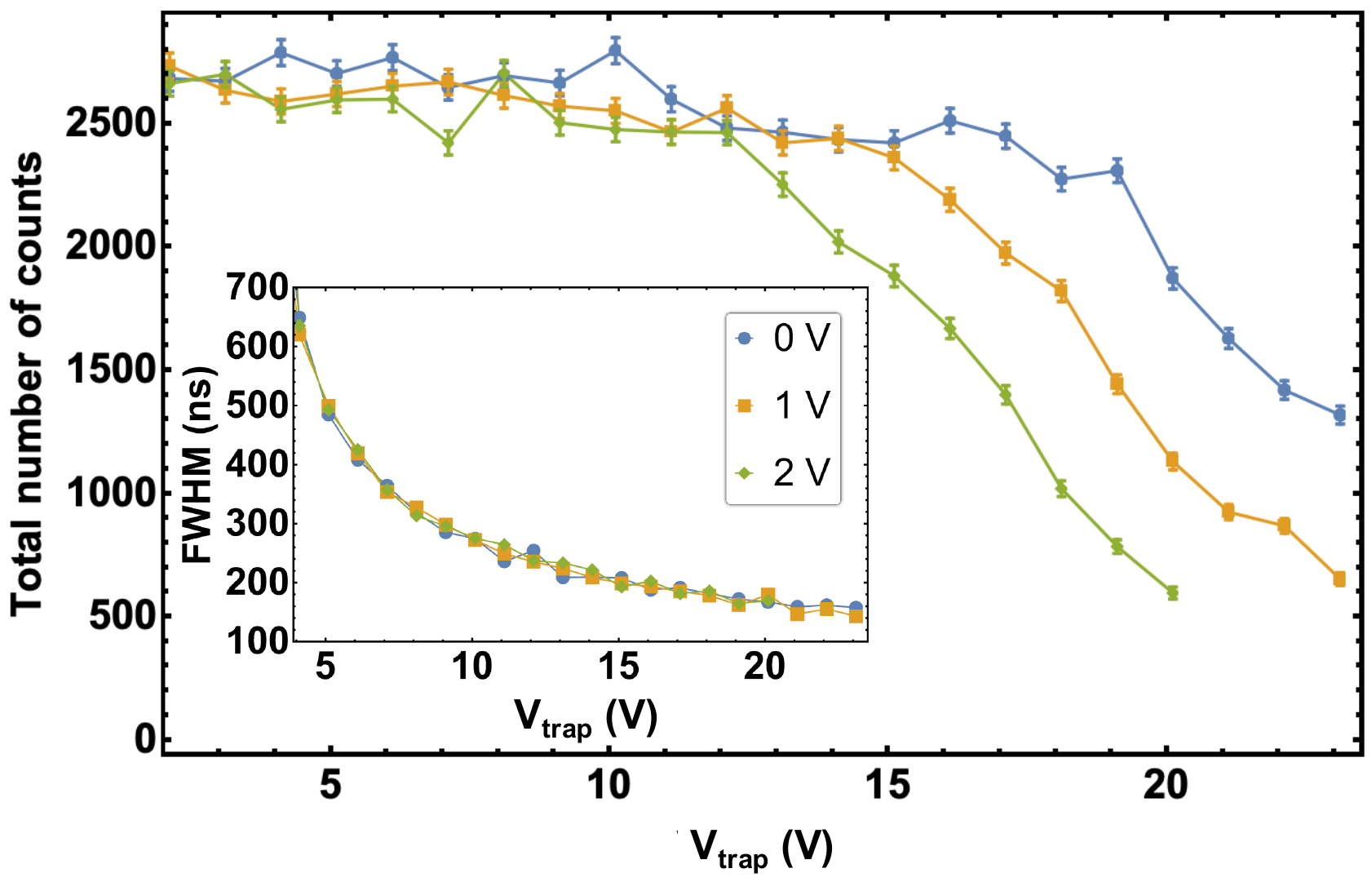}
	\caption{Number of transported ions as a function of the trap depth, ($V_{trap}$), for three potential differences between segments along the bunching region. The behavior of the FWHM of $^{39}$K$^+$ peak is given as an inset.}
\label{fig::DVbunch-Vtrap}
\end{figure} 

As indicated in Fig.~\ref{fig::DVbunch-Vtrap}, the number of transported ions is nearly the same up to a trap depth of $V_{trap} = 10$~V, where $V_{trap}$ is defined as the potential difference between electrodes I (when it is low) and J. Over that range, a small or non-existent potential slope is preferred. This slope was formed by applying a potential difference $\Delta V_{bunching}$ between adjacent segments. Fig.~\ref{fig::DVbunch-Vtrap} shows the results for $\Delta V_{bunching}$ = 0, 1, and 2~V. Furthermore, a smaller potential slope allows for a greater trap depth before ion loss occurs. The FWHM of the bunches decreases steadily with the trap depth in the same fashion, regardless of the potential slope. Next, a two-dimensional scan of electrode J and K, which sets the barrier height, was performed for $\Delta V_{bunching}$ = 0, from which a trap depth of 10~V was selected. The number of extracted ions showed little dependence on the height of the potential barrier. Similarly, the FWHM of the bunches also varied weakly but was found to be minimized if the potential on the last segment is 0~V.

\subsection{RF amplitude scan}

\begin{figure}
	\centering
	\includegraphics[width=0.95\linewidth]{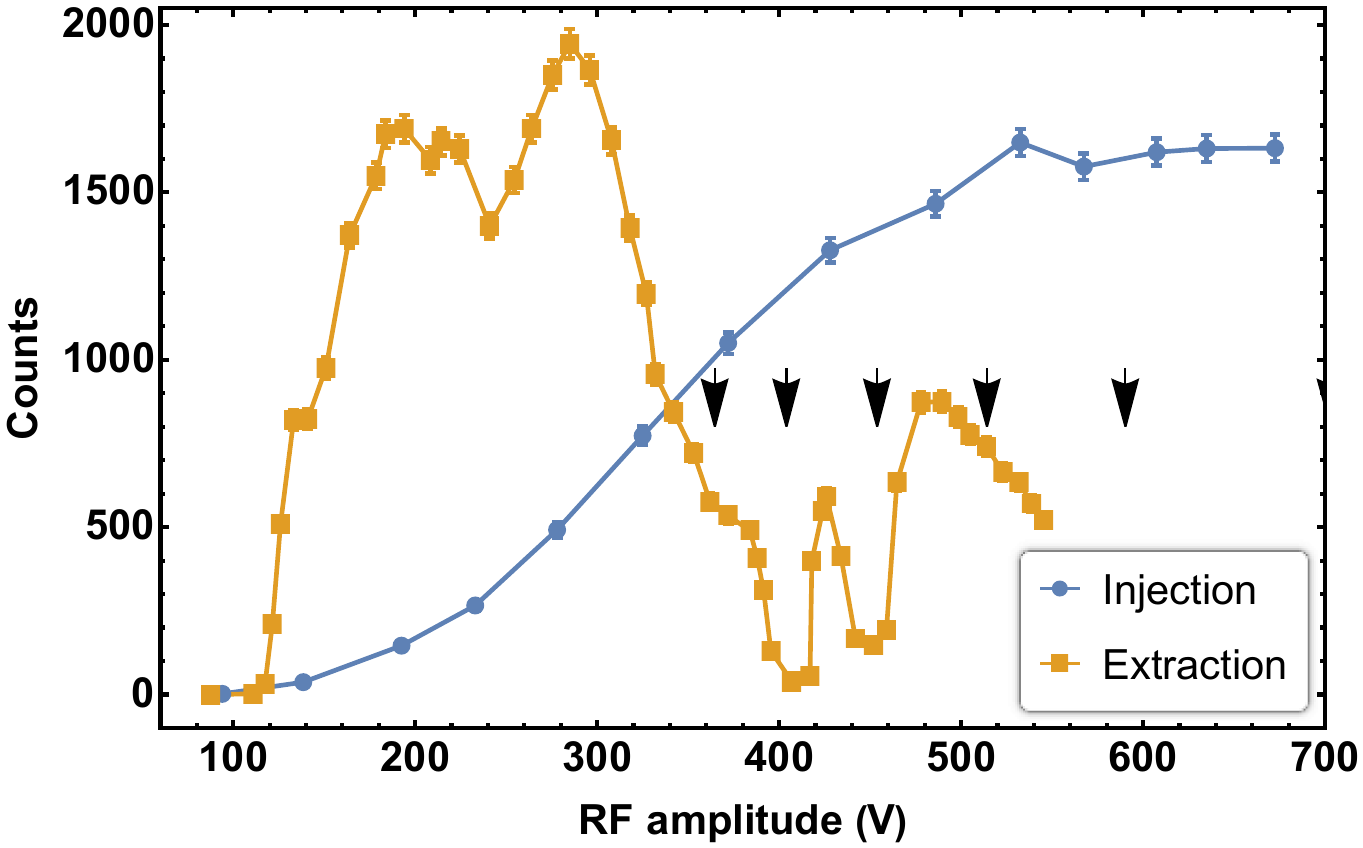}
	\caption{Number of detected potassium ions for various RF amplitudes for the injection and extraction sections. The arrows indicate extraction RF amplitudes that result from unstable ion motion \cite{Drakoudis2006}. The first arrow correspond indicates $q=0.278$ while the fifth $q=0.45$. The injection side covers lower values of $q$ that ends at $q=0.277$. Note that the extraction RF amplitude was scanned up to an amplitude of 550 V.}
	\label{fig::rfAmplitudeScan}
\end{figure} 

After the DC potentials were adjusted, the effect of the amplitude of the injection RF (covering segments A to C) and extraction RF (covering segments D to K) on the potassium ions was studied. The injection RF amplitude was first scanned with the extraction RF amplitude set at 305 V. For simplicity, the injection and extraction RF of 4.98~MHz and 3.66~MHz respectively, were not changed in those studies. As indicated in Fig.~\ref{fig::rfAmplitudeScan}, the number of recorded counts steadily increases in the RF amplitude range from 100~V until 500~V after which it plateaus. These results indicate that below $U=500$~V, the pseudopotential depth of 13~V at $r=r_0$  
is insufficient to radially confine the ion beam as it is transported in the half of the cooler-buncher covered by the injection RF. The pseudopotential depth is obtained using
\begin{equation}
V_{pseudo} = \frac{e U^2}{4m {\omega}^2  {r_o}^4} r^2  
\label{eq::pseudo}
\end{equation}
where, $e$ and $m$ are the charge and mass of the ion traversing the RFQ, $U$ is the RF amplitude, $r_o$ is the spacing between opposing rods, and $\omega$ is the angular radiofrequency, and $r$ the radial distance from the axis. 

Then, the extraction RF amplitude was scanned with the injection RF amplitude set to 635~V. Fig.~\ref{fig::rfAmplitudeScan} indicates that the number of recorded counts vary more sensitively with that RF amplitude. First, the number of counts increases rapidly when the RF amplitude changes from 110~V to 190~V. This indicates that the pseudo-potential is insufficient to radially confine the ions in the half of the cooler-buncher covered by the extraction RF if the RF amplitude is below 190~V. The lower amplitude required for the extraction section is a consequence of the smaller frequency used as Eq.~\ref{eq::pseudo} indicates. The recorded potassium counts were also observed to drop for amplitudes greater than 300~V. Increasing the RF amplitude not only creates a deeper pseudo-potential, it also increases the size of the stability parameter $q$:
\begin{equation}
q = \frac{2e U}{m \omega^2 {r_o}^2}.
\label{eq::qParameter}
\end{equation}
The ion motion in an RFQ becomes unstable when $q>0.908$ \cite{werth2009charged}, which would correspond to an RF amplitude of 1191~V, which is much larger than that studied in Fig.~\ref{fig::rfAmplitudeScan}. However, the segmentation of the quadrupoles further affects the ion motion introducing instabilities for certain values of the stability parameter $q$ \cite{Drakoudis2006}. The RF amplitudes corresponding to the first five of such unstable $q$, obtained from \cite{Drakoudis2006}, are indicated by the arrows in Fig.~\ref{fig::rfAmplitudeScan}. As can be seen, the overall drop in counts coincides with having an RF amplitude that reaches these first few unstable settings. Future operation of the cooler-buncher will maintain an RF amplitude on the ejection side with a $q$ that is below this unstable region. 
It should also be mentioned that the axial potential created by the segmentation induce a stability parameter $a_{eff}$ \cite{Drewsen2000}:
\begin{equation}
a_{eff} = \frac{4e U_{eff}}{m \omega^2 {r_o}^2},
\label{eq::aParameter}
\end{equation}
which unlike the traditional $a$ parameter, will affect the motion along the x- and y-direction in the same way. The effective potential $U_{eff}$ depends on the specific electrode geometry. Ultimately, this results in an extension of the stability region beyond the traditional $q=0.908$ albeit for a narrower range of $a_{eff}$ \cite{Drewsen2000}.  

%\begin{figure}
	%\centering
	%\includegraphics[width=0.95\linewidth]{figures/RFamplitudeMass-v2.pdf}%
	%\caption{RF amplitude bounds within which $\sim$50$\%$ of singly-charged ions get extracted from the cooler-buncher as a function of the ion's atomic mass. The lower and upper bounds are set by the dashed and solid lines respectively.}
	%\label{fig::RFamplitudeMass}
%\end{figure} 

%Based on the results of Fig.~\ref{fig::rfAmplitudeScan}, as well as Eq.~\ref{eq::pseudo} and Eq.~\ref{eq::qParameter}, we calculated the expected lower and upper bounds where at least 50$\%$ of the ions are extracted from the cooler-buncher. From Fig.~\ref{fig::rfAmplitudeScan}, these bounds correspond to 151 and 332 V RF amplitudes for a $A=39$~u singly charged ion. As indicated in Fig.~\ref{fig::RFamplitudeMass}, the cooler-buncher can be operated with an RF amplitude that leads to a transport efficiency above 50$\%$ for singly charged ions with an atomic mass as low as 8 u. These results indicate that we can efficiently transport all species we plan on measuring at St.~Benedict, the lightest of which is $^{11}$C.  

\subsection{Time-of-flight profile}

\begin{figure}
\centering
	\includegraphics[width=0.95\linewidth]{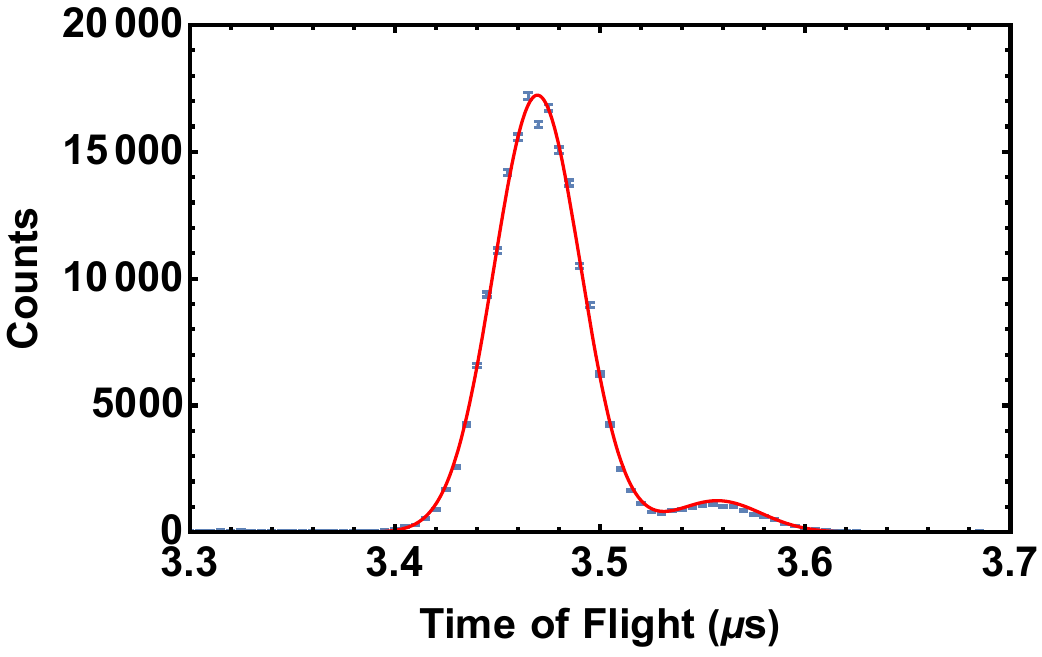}
	\caption{The sum of 10,000 consecutive bunches ejected from the cooler-buncher obtained by the MCP detector directly downstream of the chamber.  The fit, given in red, shows the two peak Gaussian fit matching the two masses coming from the ion source, $^{39}$K$^{+}$ and $^{41}$K$^{+}$, based on their separation and natural abundance.} 
	\label{fig::bunchProfile}
\end{figure}

After optimizing all settings, the sum of 10,000 bunches, whose profile is shown in Fig. \ref{fig::bunchProfile}, was obtained. It should be noted that the ejection potentials, established by having electrodes $I=260$ V and $K=-230$ V, were adjusted to minimize the FWHM without losing ions or exceeding the operational voltages of the fast switches. Two peaks with time-of-flight of 3.47 and 3.56 $\mu$s can be clearly seen and have been identified as the two isotopes coming from the source, $^{39}$K$^{+}$ and $^{41}$K$^{+}$. In red we overlay a fit of two Gaussians with a height ratio equal to the abundance ratio of $^{41}$K/$^{39}$K = 0.072 and the separation fixed by their mass ratio. This fit reproduces the observed spectra quite well and gives a full width half maximum of 49.8(3) ns for the bunches. This is smaller than the results presented in Fig.~\ref{fig::DVbunch-Vtrap} because then, the ions were release by only lowering the potential on electrode K while here, the simultaneous raising of the potential on electrode I results in a temporal compression of the bunch. However, this can come at the detriment of an increased energy spread. A determination of the energy spread will be subject to future studies once the cooler-buncher is being recommissioned on-line. The optimized settings used to obtain this bunch are outlined in Table \ref{tab::optimizedVoltages} and Table \ref{tab::optimizedParameters}.

\begin{table}
	\centering
	\caption{Optimized DC voltages for the RFQ cooler-buncher.}
	\begin{tabular}{ l r }
		\hline
		\hline
		Electrode & Optimized Voltage (V) \\
		\hline 
		Source Voltage  & 200 \\
		Source Anode  & 193.5 \\
		Injection Lens & 0/300 *\\
		Injection Hyperboloid  & -1800 \\
		Injection Cone & -1000 \\
		RFQ Electrodes A, B, D & 0 \\
		RFQ Electrodes C, E-H & -3.6 \\
		RFQ Electrode I & -3.6/260 **\\
		RFQ Electrode J & -11.6 \\
		RFQ Electrode K  & 0/-230 **\\
		Ejection Cone & -230 \\
		Ejection Hyperboloid & -6000 \\
		Ejection Lens & -6000 \\
		\hline 
		\hline
	\end{tabular}
	
	\begin{tabular}{l}
		*injection/blocking voltage\\
		**accumulation/ejection voltage
	\end{tabular}
	\label{tab::optimizedVoltages}
\end{table}

\begin{table}
	\centering
	\caption{Optimized RF, timing, and pressure settings for the RFQ cooler-buncher.}
	\begin{tabular}{ l r }
		\hline
		\hline
		Parameter & Optimized Setting \\
		\hline 
		Injection RF  & 4.98 MHz \\
		Injection RF Amplitude  & 635 V \\
		Ejection RF  & 3.66 MHz \\
		Ejection RF Amplitude  & 305 V \\
		Cycle Time & 50 ms \\
		Accumulation Time & 10 $\mu$s \\
		Holding Time & 49 ms \\
		RFQ Helium Pressure & 0.063 Torr \\	
		\hline 
		\hline
	\end{tabular}
	\label{tab::optimizedParameters}
\end{table} 

\subsection{Bunching Efficiency}

After optimizing the cooler-buncher, we determined its bunching efficiency. This was accomplished by taking the ratio between the number of ions transmitted through the RFQ in bunching mode and in shoot-through mode. In both cases a 10 $\mu$s stream of ions was produced by lowering the potential on the injection lens electrode for that duration.
A total of 17,235 counts for 10,000 scans were counted in shoot-through mode over the $\approx$50 ms recording window that included the complete count distribution. Recording for the same duration without beam resulted in 257 background counts. Finally, the total number of counts in bunching mode was found to be 15,825 ions. Based on this measurement, the bunching efficiency was found to be 93(1)$\%$.

%The bunching efficiency of the RFQ was determined by optimizing the injection optics, DC gradient, RF amplitude, and buffer gas pressures while varying the extraction hyperbola and cone electrodes. Scans were taken with a continuous beam—the high voltage switches for electrodes I and K set to their low values such that ions are not stopped in the bunching section—and compared to scans taken in bunching mode. This data was analyzed in the same manner as outlined in section 4.2 and results were plotted as a contour plot of total counts over the parameter space (see figure \ref{fig::bunchEfficiency}). While optimal settings for the extraction optics were different for each of the two cases, at their respective maxima the same number of ions were transmitted through the RFQ. It was determined that 100\% of the ions which make it through the RFQ without bunching, are also transported when the ion beam is being bunched.

%\begin{figure}
	%\includegraphics[width=\linewidth]{figures/bunching_efficiency.pdf}
	%\caption[Bunching efficiency of cooler-buncher.]{Contour plots showing total counts collected at the MCP as a function of voltage applied to the extraction hyperbola and cone electrodes. The top figure shows the maximum for DC beam while the lower figure shows the maximum for a bunched beam.} 
	%\label{fig::bunchEfficiency}
%\end{figure} 

\subsection{Trapping lifetime}

\begin{figure}
\centering
	\includegraphics[width=0.95\linewidth]{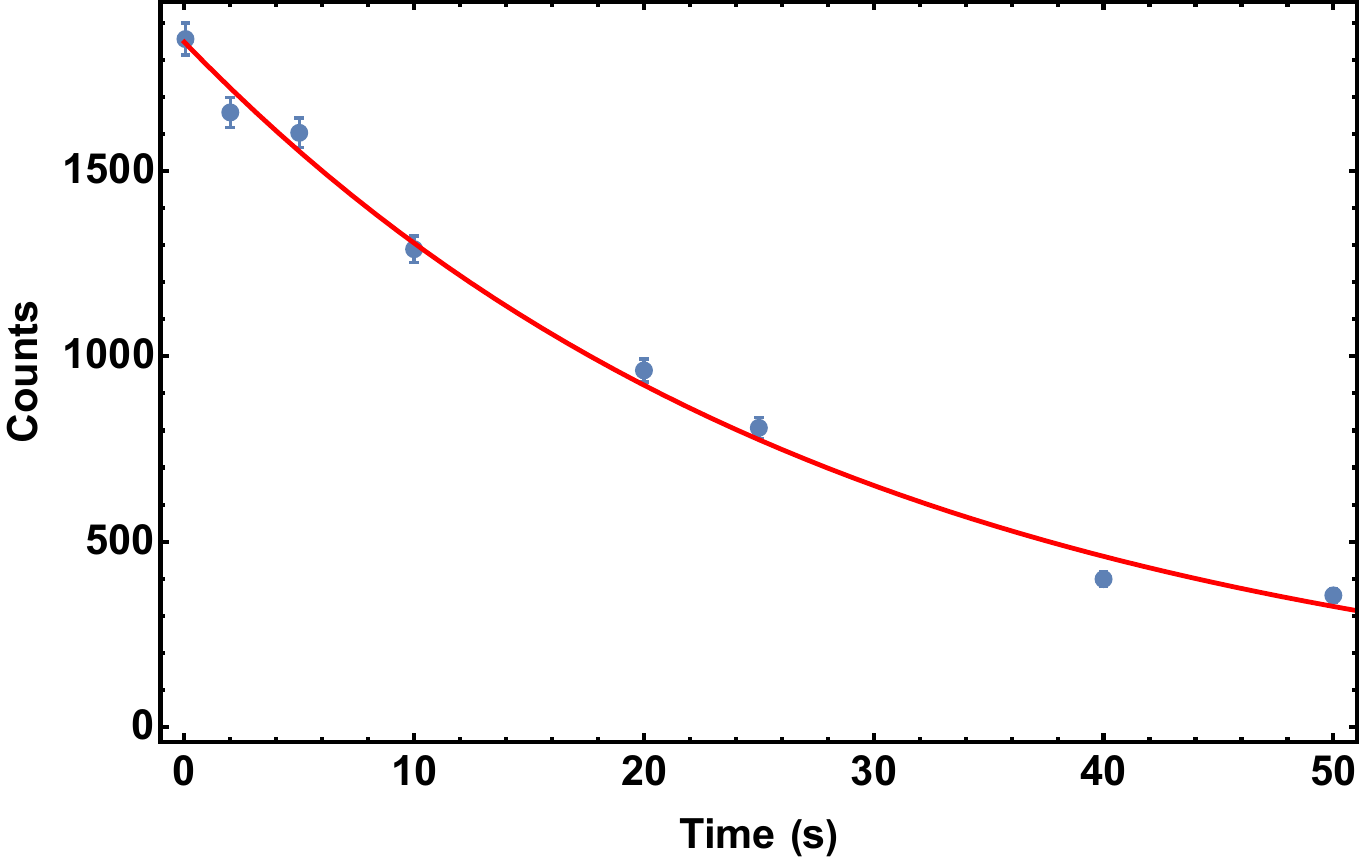}
	\caption{Number of recorded potassium ions extracted from the cooler-buncher as a function of the time spent in the trap with exponential fit denoting a 20.0(5) s trapping half-life.} 
	\label{fig::lifetime}
\end{figure}

Under realistic conditions, a group of ions cannot be confined indefinitely and some ions will be lost over time by various mechanisms, including charge exchange or collisions with molecular contaminants. As a result, the number of ions inside the bunching section of the RFQ should decrease exponentially over time. This effect was studied for the St.~Benedict cooler-buncher by varying $t_{hold}$ from 49 ms to 49.999 s. Each time, the duration of the cycle was adjusted to be 1 ms longer than the holding time. Each data point in Fig.~\ref{fig::lifetime} is the total number of counts of 2,000 bunches. Because all data points greater than 49 ms take over an hour to complete, we interleaved each of these measurements with a 49 ms holding time measurement to monitor temporal stability in the number of counts. An exponential fit of the number of recorded counts as a function of time spent in the trap yields a trapping half-life of 20.0(5) s for potassium ions. 

\section{Conclusion and Outlook}

The St.~Benedict RFQ cooler-buncher has been commissioned using an off-line potassium ion source. %demonstrating 80(3)$\%$ injection efficiency of the beam from the ion source and the complete transport of the ion beam entering the device in shoot-through mode. 
The various electrode potentials and the RF amplitude have been optimized resulting in the extraction of ion bunches with a FWHM of 50 ns at its exit and a bunching efficiency of 93(1)$\%$ when using a 50 ms trapping cycle. The trapping half-life of the potassium ions has been measured to be 20.0(5) s. This is a sufficiently long duration compared to the foreseen duty cycle of 1 Hz or more, at which the cooler-buncher will operate. These commissioning results not only provide starting parameters and expectations for the commissioning of the cooler-buncher once installed on-line but also showed that it can efficiently extract narrow ion bunches. Note that the measurements presented were done with only a few ions per bunch. As the number of incoming ions increases, the effect of space charge will modify the TOF distribution, and at sufficiently high beam rate it will affect the transport and bunching efficiency. Once installed on-line, further studies including the effect of space charge, the transport and lifetime of different masses, as well as measurements of the energy spread of the bunches, and the purity of the gas are planned.   

\section*{Acknowledgment}

This work was carried out with the support of the National Science Foundation under grants PHY-1725711, PHY-2310059, PHY-2050527, the University of Notre Dame, as well as the US Department of Energy, Office of Science, Office of Nuclear Physics under Contract No. DE-AC02-06CH11357 (ANL). A.A. Valverde acknowledges the support of the Natural Sciences and Engineering Research Council of Canada under grant No. SAPPJ-2018-00028.

\end{document}